\documentclass[12pt,aaspp4]{aastex}
\newcommand{\rsun}{R$_{\odot}$}
\newcommand{\ks}{km~s$^{-1}$}

\begin{document}

\title{Identification of the Coronal Sources of the Fast Solar Wind} 
\author{S. Giordano, E. Antonucci}
\affil{Osservatorio Astronomico di Torino, 
	Strada Osservatorio 20, Pino Torinese 10025, Italy}
\affil{giordano@to.astro.it}
\author{G. Noci, M. Romoli}
\affil{Universit\`a di Firenze, Dipartimento di Astronomia e Scienza dello Spazio, 
	L.go E. Fermi 5, Firenze 50125, Italy}
\and
\author{J.L. Kohl}
\affil{Harvard-Smithsonian Center for Astrophysics, 
	60 Garden Street, Cambridge, MA 02138}

\begin{abstract}

The present spectroscopic study of the ultraviolet coronal emission in a
polar hole, detected on April 6--9, 1996 with the Ultraviolet Coronagraph 
Spectrometer aboard the SOHO spacecraft, identifies the inter--plume lanes 
and background coronal hole regions as the channels where the fast solar wind 
is preferentially accelerated. In inter--plume lanes, at heliocentric 
distance  1.7~\rsun, the corona expands at a rate between 105~\ks\, 
and 150~\ks, that is, much faster than in plumes where the outflow 
velocity is between 0~\ks\, and 65~\ks. 
The wind velocity is inferred from the Doppler dimming of the 
O~VI $\lambda\lambda$~1032, 1037~\AA\, lines, within a range of values, 
whose lower and upper limit corresponds to anisotropic and isotropic 
velocity distribution of the oxygen coronal ions, respectively. 

\end{abstract}

\keywords{solar wind -- Sun: corona -- Sun: UV radiation}

\newpage

\section{INTRODUCTION}

One of the primary objectives of the SOHO mission is the identification 
of the source and acceleration mechanisms of the fast solar wind. 
Although polar coronal holes and regions with open magnetic field configuration 
were recognized long ago to be at the origin of the fast wind, the paucity of 
direct coronal hole observations from space in the long interval between the 
Skylab and SOHO missions prevented in the last two decades a real progress in 
studying how the fine structure of coronal holes regulates coronal expansion. 
Polar plumes, that are the most prominent features in polar coronal holes, 
and inter--plume regions have both been proposed to play an important role in 
the generation of the high--speed wind 
(e.g., Ahmad and Withbroe, 1977; Wang, 1994). 
SOHO observations have confirmed that plumes are denser, cooler and less dynamic 
structures than the surrounding regions \citep{dos97,noc97,wil98}.
Nevertheless they are site of quasi--periodic compressional waves \citep{def98}, 
identified as slow magnetosonic ones that however can carry only a 
fraction ($\sim$2$\times$10$^3$~erg~cm$^{-2}$~s$^{-1}$) of the energy required 
to accelerate the fast solar wind, $\sim$10$^5$~erg~cm$^{-2}$~s$^{-1}$ 
\citep{ofm99}.
The Ultraviolet Coronagraph Spectrometer (UVCS) onboard SOHO, is the first 
instrument that can allow us to determine the outflow velocity of the wind 
plasma in the outer corona and therefore to distinguish between the dynamic 
conditions of plumes and surrounding regions, at the height where the wind 
velocity has become significant.
The results of such an analysis performed on the basis of the Doppler 
dimming of the O~VI $\lambda\lambda$~1032, 1037~\AA, detected in the period 
April 6--9, 1996, are reported in the present paper. 
Preliminary results of this study are found in the Ph.D. thesis by \citet{gio98}. 

\section{OBSERVATION OF THE POLAR CORONAL HOLE}

In the first detailed observation of the ultraviolet emission of a coronal hole 
above 1.5~\rsun, performed at solar minimum with UVCS during April 6--9, 1996,
the polar region was scanned over an interval of 72~hours, starting 
on April 6, 1996 at 07:22~UT.
The instantaneous field of view (29~arcmin\,x\,14~arcsec), centered on the North 
pole, was moved by 14~arcsecond steps in the radial direction, thus ensuring a 
continuous coverage of the corona between 1.45 to 2.48~\rsun.
For each spatial element, 14"\,x\,14", the two O~VI lines, at 
$\lambda$~1031.91~\AA\, and $\lambda$~1037.61~\AA\, were detected with spectral 
resolution of 0.2~\AA, and integration time of 3600~seconds. 
The raw data are calibrated according to the standard procedure \citep{gar96}. 
The O~VI line profiles are then fitted with a function resulting from the 
convolution of a gaussian function, for the solar spectral profile, with the 
Voigt curve describing the instrumental broadening, and a function accounting 
for the width of the slit \citep{gio98}. 
The best fit is obtained by adjusting as free parameters standard deviation, 
$\sigma_{\lambda}$, mean wavelength, $\lambda_o$, and peak, $I(\lambda_o)$, of 
the  solar profile. 
The observed line intensity is then the integral 
$I_{tot} = \sqrt{2\pi}~I(\lambda_o)~\sigma_{\lambda}$. 

In the O~VI images of the polar hole observed on April 6--9, plumes are clearly 
identified at least up to 2~\rsun. 
Four main plumes are present within $\pm$14$^{\circ}$ from the North pole, 
as the O~VI~1032 image shows in Figure~1. 
They appear as bright broad features, dimming with heliodistance. 
Since the observation time to scan the corona out to 2~\rsun\, is 1.4 days, 
plumes are either fairly stable or they tend to form again in almost the same 
position. 
In 1.4 days the displacement of the plumes due to solar rotation is negligible. 
Their width at 1.7~\rsun\, is roughly 5$\times$10$^9$--10$^{10}$~cm. 
Outside the central region, $\ge\vert\pm 14^{\circ}\vert$, plumes are weaker and 
fewer as shown by the intensity along the heliocentric circumference with 
radius 1.7~\rsun\, (Figure~1).

Aim of the analysis is to determine the solar wind velocity in plumes and 
surrounding regions, including inter--plume lanes and darker areas of lower plume 
population, outside $\pm$14$^{\circ}$, that is, background coronal hole regions, 
with the intent of identifying the source of the high--speed wind. 
Therefore the different regions are studied at a height 1.7~\rsun, where plumes 
are still well identified and the wind has acquired on the average a sufficiently 
high velocity, $\geq$~100~\ks\, \citep{str93,ant97a,ant97b,koh97a,koh98,gio99}.

The plume emission is averaged over the brightest peaks within  
$\pm$14$^{\circ}$ identified as dark segments in Figure~2, and the lane emission 
is averaged over the dimmest regions (dashed segments in Figure~2). 
The emission of dark background regions is averaged outside the interval 
$\pm$14$^{\circ}$.
The background average heliodistance, 1.82~\rsun, is higher than that of plumes 
and inter--plume lanes, 1.72~\rsun. 

\section{WIND VELOCITY IN PLUMES AND LANES} 

The solar wind velocity relative to the three different regimes that have been 
identified, plume, inter--plume lanes and dark background, is then measured by 
determining the outflow velocity of the oxygen ions through the ratio of the 
Doppler dimmed O~VI~$\lambda$~1032 and $\lambda$~1037~\AA\, resonance lines 
\citep{noc87,dod98,li98}.
 
In the outer corona these lines are emitted both via collisional excitation 
of the coronal ions and via resonant scattering of photons coming from the 
transition region. 
The second process is of increasing importance as the corona becomes more 
rarefied. 
In the frame of reference of the expanding solar wind the relative wavelength 
shift of incident photons and coronal absorbing profiles, causes a dimming of 
the  resonant emission that is a function of the outflow velocity \citep{bec74}.
The wind velocity can then be determined by the intersections of the observed 
O~VI intensity ratios with the emissivity ratios curves calculated with the code 
by \citet{dod98}, for the coronal conditions observed, or inferred, at 1.7~\rsun.
In a spherical symmetric corona, the ratios of the emissivities on the plane of 
the sky closely approximates the line intensity ratios \citep{noc87}.

The electron temperature, $T_{e}$, is deduced from the SOHO observations that 
indicate values that remain below 1$\times$10$^6$~K in a coronal hole and 
decrease above 1.15~\rsun\, \citep{dav98}. 
A rare measurement in a bright plume, at 1.6~\rsun, yields a 4$\times$10$^5$~K 
temperature \citep{wil98}. Higher coronal temperature values are determined on 
the basis of in--situ charge state ionization measurements, performed with 
SWICS/Ulysses, which are however derived in the assumption of slow wind \citep{ko97}. 
Here, we assume the temperature roughly equal to 3$\times$10$^5$~K at 1.7~\rsun\, 
for both plumes and surrounding regions. The wind velocity results do not change 
in any significant way if a  higher temperature, e.g. 1$\times$10$^6$~K, is 
assumed outside plumes.
The plume electron density is derived from white light observations performed 
during minimum solar activity in 1996 \citep{gua99}.
Since the inter--plume and background region density is not explicitly given 
in that paper, it is assumed to be equal to the quiet coronal hole (Table~1). 
The ratio of the plume and inter--plume density is consistent with the values 
published by \citet{cra99}.

The O~VI coronal absorbing profiles along the line--of--sight are  directly 
measured with UVCS.  The width of the ion velocity distribution along the
line--of--sight, 
$ {v_{1/e}}\, = \sqrt{2} \, {c \over {\lambda_{o}}} \, \sigma_{\lambda}$, 
is equivalent to the kinetic temperature, 
${T_{k}} \, = \, {{m_{i}} \over {2 k_{B}}} \, {v^{2}_{1/e}}$, 
where $k_{B}$ is the Boltzmann constant and $m_{i}$ is the ion mass.
Table~2 reports the observed intensity ratios, $\rho$, of the OVI~1037 to the 
OVI~1032 line and the observed kinetic temperature of the O~VI ions along the 
line--of--sight, $T_{k}$, together with  the average heliodistance of the plume, 
inter--plume and background regions.
The observed $T_{k}$ values (Table~2) confirm that the width of the ion velocity 
distribution is broader in lanes and background regions than in plumes
\citep{ant97b, noc97}.

In the computation of the O~VI emissivity the ion velocity 
distribution is  considered to be  bi--maxwellian.
Since the broadening of the O~VI lines  in regions of open magnetic 
field lines is much larger than expected on the basis of ion--proton and 
proton--electron thermal balance \citep{ant97a, koh97a, koh97b, noc97}, 
it is reasonable to assume the width of the radial distribution (perpendicular 
to the line--of--sight) as a variable between two extreme cases, that is, 
oxygen--electron thermal balance along the radial direction and isotropic 
distribution, respectively. 
Furthermore, the existence of a degree of anisotropy in the oxygen velocity 
distribution is proven at least above 1.8~\rsun\, 
\citep{koh97a, koh98, ant99, cra99}.
The emissivity ratios are then computed both for the radial velocity distribution 
characterized by $T_{k,r} = T_{e}$, (i.e. maximum of anisotropy, corresponding to 
oxygen ions and electrons in thermal equilibrium along the radial direction) in 
Figure~3, and by $T_{k,r} = T_{k}$, (i.e. isotropic velocity distribution) 
in Figure~4.
Calculated and observed emissivity ratio of the O~VI lines are compared to derive 
the outflow velocity.

\section{OUTFLOW VELOCITY RESULTS}

The expansion velocity, derived from the intersection of the emissivity ratio 
curve with the observed O~VI ratio, in Figure~3 and 4, has not a  unique value 
because of the lack of information  on the radial ion velocity distribution, 
but it falls within a higher limit derived  for isotropic oxygen 
velocity distribution and a lower value for the anisotropic case. 

It is immediately obvious from the inspection of Figure~3 and 4, that lanes 
between plumes and dimmer background regions outside $\pm14^{\circ}$ are the 
privileged sites for the fast solar wind acceleration. 
In these regions, at 1.7~\rsun, the fast solar wind has already reached speeds 
above 100~\ks, while the plasma in plumes is either expanding slowly or remains 
almost static. In lanes the outflow velocity is between 105~\ks\, and 150~\ks, 
and in darker regions between 110~\ks\, and 180~\ks\, for anisotropic and isotropic 
ion velocity distribution, respectively (Table~3). 
These values  are of the order of  the average outflow velocity detected  
from the O~VI line  emission  integrated along the instantaneous field--of--view,  
at this height, which is between 100~\ks\, and 125~\ks\, \citep{gio99}.
The tendency to measure higher velocity in the darker background regions might 
reflect the fact that they are located on the average at higher heliodistances,
where the wind is faster. 
The errors of the outflow velocity derived for isotropy are larger because 
in this case the derivative of the emissivity ratio is smaller than in the case of 
anisotropy, as shown in Figure~3 and 4. 
No significant outflow velocity can be found in plumes  for isotropic velocity 
distribution. 
However if the distribution of the oxygen ion velocity deviates from isotropy, 
it is possible to detect a slow expansion at 65~\ks, much lower than that  found 
for inter--plume lanes and background regions.

\section{CONCLUSIONS}

As a result of the spectroscopic analysis of the ultraviolet emission from a
solar minimum polar hole in the outer corona, we can therefore conclude that 
there is strong evidence that the fast solar wind is preferentially accelerated 
in inter--plume lanes and in the darker background regions of a polar coronal hole. 
It is interesting to note that these are also the regions where the O~VI line 
broadening is enhanced relative to that observed in  plumes \citep{ant97b, noc97}. 

According to the UVCS results (e.g., Kohl et al., 1997a), the wind acceleration 
is higher where line profiles are broader. 
This fact has led to interpret the excess line broadening in terms of preferential 
heating of the oxygen ions related to preferential acceleration of the solar wind.
The correlation of higher outflow velocity and broader line profiles outside 
the plumes is consistent with this interpretation. This does not contradict
the model by \citet{wan94}, that invokes higher heating at the base of plumes,
since the enhanced broadening of the oxygen lines in inter--plume lanes, evidence 
for preferential heating of the oxygen ions, is probably related to wave-particle 
interaction occurring higher up in the corona \citep{koh98,cra99}.

The present results, that identify the regions surrounding plumes as the 
dominant sources of the high--speed solar wind, together with those recently 
obtained by \citet{has99} and \citet{pet99}, who observed blue shifts at the base 
of polar coronal holes, suggesting that solar wind is emanating from regions along 
the boundaries of magnetic network cells, are finally indicating how to track back 
to the coronal base the fast solar wind velocity field lines.
 
\acknowledgments
 The authors wish to acknowledge the financial support of the Italian Space 
 Agency and NASA.

\newpage

\begin{table}[htb]
 \begin{center}
    \caption{CORONAL HOLE PHYSICAL PARAMETERS} 
    \leavevmode
    \begin{tabular}[h]{lccc}
      \hline \\[-5pt]
      		 	& Distance & $N_{e}$		   &  $T_{e}$  	    \\ 
       		 	& (\rsun)  & (10$^{5}$\,cm$^{-3}$) &  (10$^{5}$\,K) \\ 
      \hline \\[-5pt]
	Plume 		& 1.72	   & 5.1  		   & 3.0 	    \\
	Lane 	 	& 1.72	   & 3.6  		   & 3.0 	    \\
	Background 	& 1.82	   & 2.6  		   & 3.0 	    \\
	 \hline \\
      \end{tabular}
    \label{tab:tab1}
  \end{center}
\end{table}

\begin{table}[htb]
  \begin{center}
    \caption{O~VI LINE RATIO AND KINETIC TEMPERATURE} 
    \leavevmode
    \begin{tabular}[h]{lccc}
      \hline \\[-5pt]
      			& Distance   & $\rho$  	       &  $T_{k}$	\\ 
       			&   (\rsun)  &   	       &  ($10^{7}$~K)	\\ [+5pt]
      \hline \\[-5pt]
	Plume 	 	&	1.72 & 0.34 $\pm$ 0.03 & 2.7 $\pm$ 0.3 	\\
	Lane 	 	&	1.72 & 0.42 $\pm$ 0.05 & 3.2 $\pm$ 0.5 	\\
	Background 	&	1.82 & 0.43 $\pm$ 0.04 & 4.5 $\pm$ 0.6 	\\
	 \hline \\
      \end{tabular}
    \label{tab:tab2}
 \end{center}
\end{table}

\begin{table}[htb]
  \begin{center}
    \caption{MEASURED OUTFLOW VELOCITY ($^*$)} 
    \leavevmode
    \begin{tabular}[h]{lcc}
      \hline \\[-5pt]
      			& $T_{k,r}=T_{e}$	 & $T_{k,r}=T_{k}$ 	\\ 
      \hline \\[-5pt]
	Plume 	 	& \,\, 65 (-25,+15) 	 & $\sim$~0  		\\
	Lane 	 	&	   105 (-17,+11) &    150 (-57,+42) 	\\
	Background	&	   110 (-10,+10) &    180 (-46,+39) 	\\
      \hline \\
      \end{tabular}
    \tablenotetext{*}{in units (\ks)}
    \label{tab:tab3}
  \end{center}
\end{table}

\begin{figure}
\epsscale{1.0} \plotone{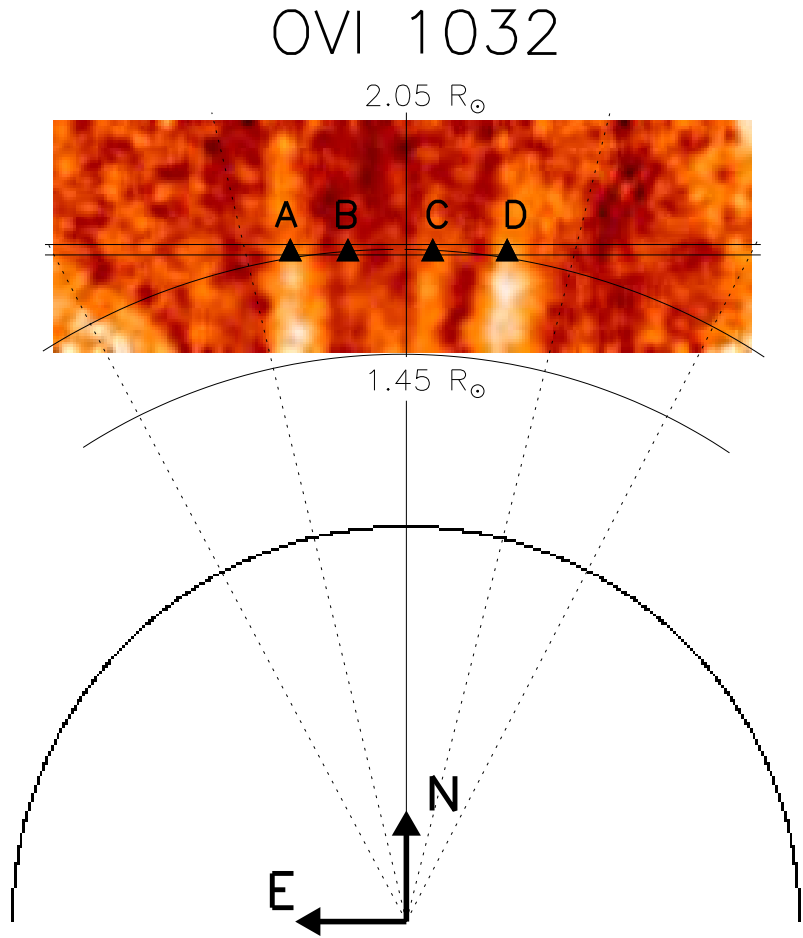}
\caption{OVI 1032 image of the polar coronal hole observed in the period 
April 6--9,1996. The field of view considered for the analysis,
at 1.7~\rsun, is delimited by continuous horizontal lines. 
The background coronal hole regions are separated from the central plumes 
by the radial dashed lines. The four brightest plumes extending up 
to 1.7~\rsun\, are identified by letters and their centroid is marked by 
solid triangles. 
\label{fig:fig1}}
\end{figure}

\begin{figure}
\epsscale{1.0} \plotone{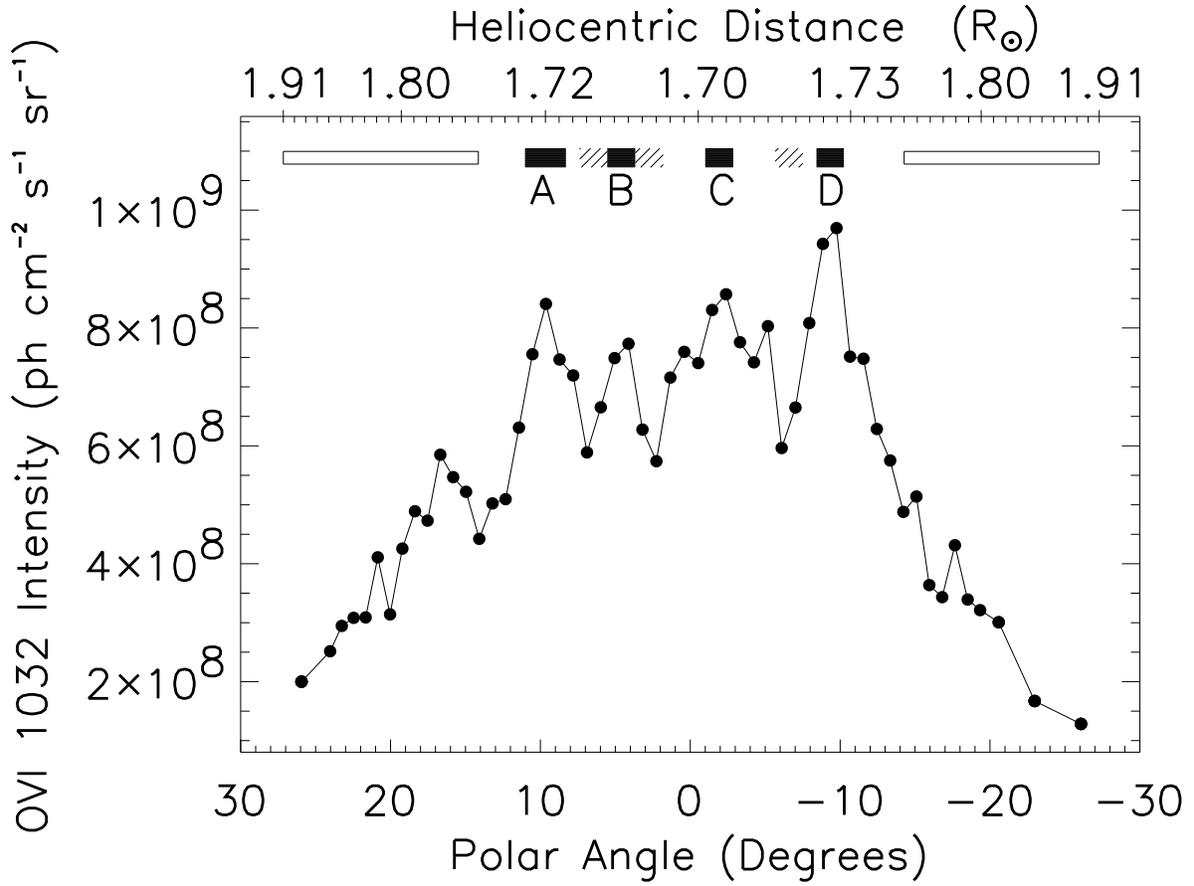}
\caption{O~VI 1032 intensity along the field of view at 
1.7~\rsun; dark segments identify plumes, dashed segments inter--plume 
lanes and open rectangles the dark background. The intensity is 
obtained by summing two contiguous observations, 
with mirror pointed at 1.69~\rsun\, and at 1.71~\rsun. 
\label{fig:fig2}}
\end{figure}

\begin{figure}
\epsscale{1.0} \plotone{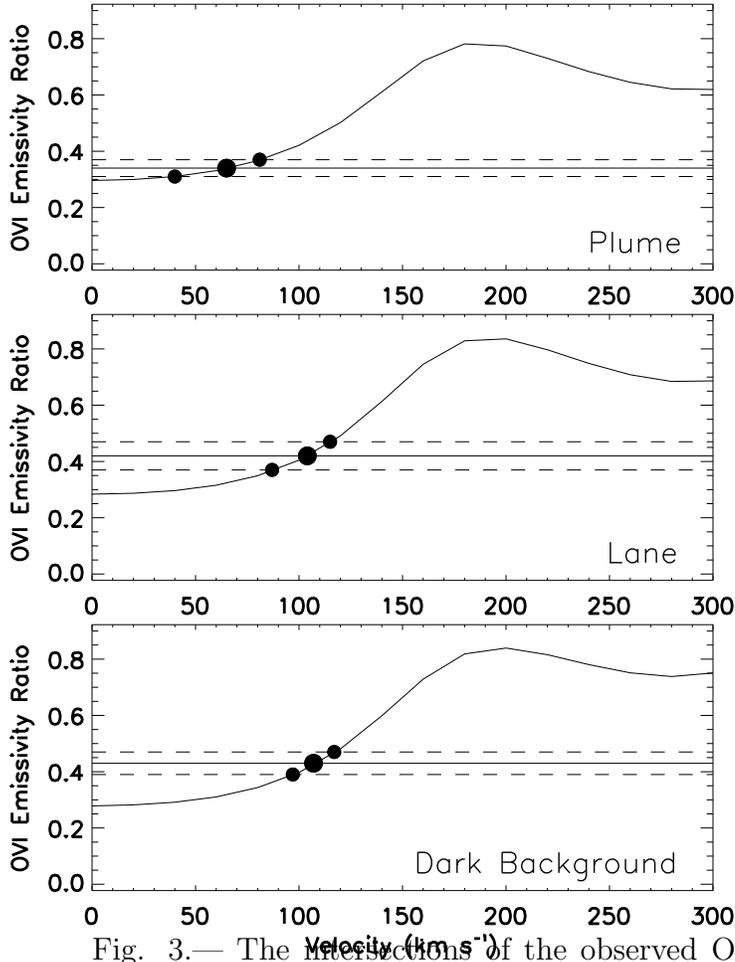}
\caption{The intersections of the observed O~VI line ratios, continuous 
horizontal lines, and calculated emissivity for  anisotropic oxygen velocity 
distribution, continuous curves, determine the outflow velocity of the 
high--speed solar wind at 1.7~\rsun, in  the  different coronal hole regions. 
The intersections with the dashed horizontal lines determine  the  statistical 
uncertainty of the measurements. 
\label{fig:fig3}}
\end{figure}

\begin{figure}
\epsscale{1.0} \plotone{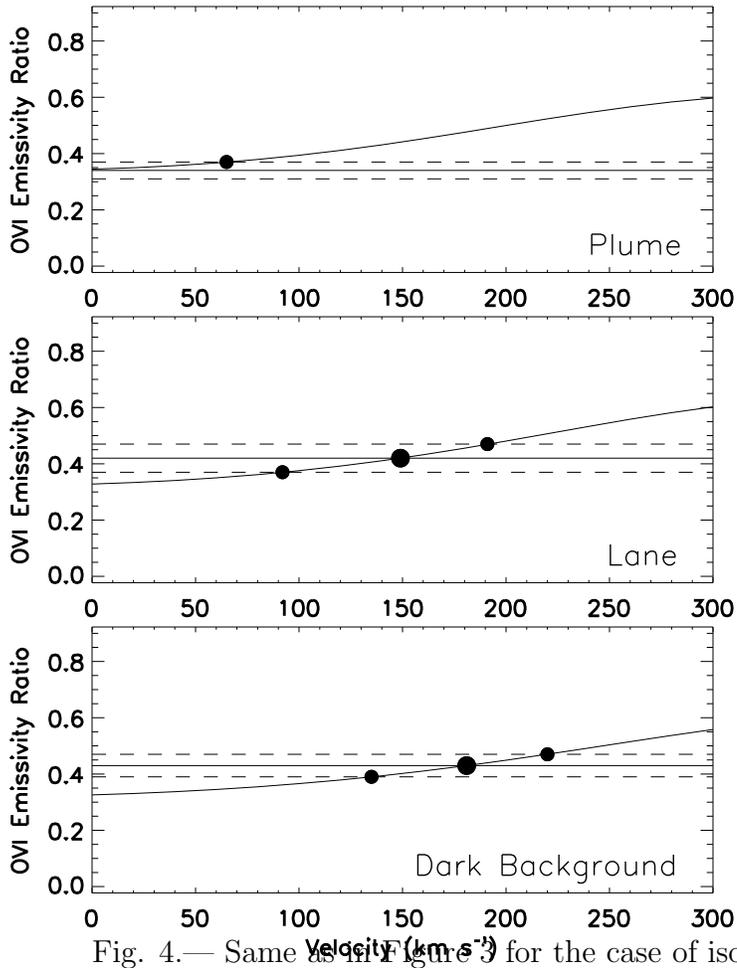}
\caption{Same as in Figure~3 for the case of isotropic oxygen velocity 
distribution. 
\label{fig:fig4}}
\end{figure}


\newpage

\begin{thebibliography}{}

\bibitem[\protect\astroncite{Ahmad and Withbroe}{1977}]{ahm77}
Ahmad, I.A., and Withbroe, G.L.: 1977,
{\it Solar Physics},
53, 397.

\bibitem[\protect\astroncite{Antonucci et al.}{1997a}]{ant97a}
Antonucci, E.
et al.: 1997a,
{\it Fifth SOHO Workshop ESA SP--404}, 175. 

\bibitem[\protect\astroncite{Antonucci et al.}{1997b}]{ant97b}
Antonucci, E.
et al.: 1997b,
{\it Astronomical Society of the Pacific Conference Series}, 118, 273.

\bibitem[\protect\astroncite{Antonucci, Giordano and Dodero}{1999}]{ant99}
Antonucci, E., Giordano, S. and  Dodero, M.A.: 1999,
et al.: 1999,
{\it  Adv. Space Res.}, in press.

\bibitem[\protect\astroncite{Beckers and Chipman}{1974}]{bec74}
Beckers, J.M., and Chipman, E.: 1974, 
{\it \solphys}, 34, 151. 

\bibitem[\protect\astroncite{Cranmer et al.}{1999}]{cra99} 
Cranmer, S.R.
et al.: 1999,
{\it \apj}, 511, 481.

\bibitem[\protect\astroncite{David et al.}{1998}]{dav98} 
David, C.
et al.: 1998, 
{\it \aap}, 336, L90.

\bibitem[\protect\astroncite{DeForest and Gurman}{1998}]{def98} 
DeForest, C.E., and Gurman, J.B.: 1998
{\it \apj}, 501, L217.

\bibitem[\protect\astroncite{Dodero et al.}{1998}]{dod98} 
Dodero, M.A., Antonucci, E., Giordano, S. and Martin, R.: 1998,
{\it \solphys}, 183, 77.

\bibitem[\protect\astroncite{Doschek et al.}{1997}]{dos97} 
Doschek, G.A.
et al.: 1997,
{\it \apj}, 477, L119. 

\bibitem[\protect\citeauthoryear{Gardner et al.}{1996}]{gar96} 
Gardner, L.D. 
et al.: 1996,
{\it Proc. SPIE}, 2831, 2-24.

\bibitem[\protect\astroncite{Giordano}{1998}]{gio98}
Giordano, S.: 1998,
{\it PhD Thesis, University of Torino}.

\bibitem[\protect\astroncite{Giordano, Antonucci and Dodero}{1999}]{gio99}
Giordano, S., Antonucci, E. and Dodero,M.A.: 1999,
{\it Adv.  Space Res.}, in press.

\bibitem[\protect\astroncite{Guhathakurta et al.}{1999}]{gua99} 
Guhathakurta, M.
et al.: 1999,
{\it \jgr}, 104, A5, 9801.

\bibitem[\protect\astroncite{Hassler et al.}{1999}]{has99}
Hassler, D.M.
et al.: 1999,
{\it Science} , 283, 810

\bibitem[\protect\astroncite{Hyder  and Lites}{1970}]{hyd70}
Hyder, C.L. and Lites, B.W.: 1970, 
{\it \solphys}, 14, 147.

\bibitem[\protect\astroncite{Ko et al.}{1997}]{ko97} 
Ko, Y.-K.
et al.: 1997,
{\it \solphys}, 171, 345. \\

\bibitem[\protect\astroncite{Kohl et al.}{1997a}]{koh97a} 
Kohl, J.L.
et al.: 1997,
{\it \solphys}, 175, 613.

\bibitem[\protect\astroncite{Kohl et al.}{1997b}]{koh97b} 
Kohl, J.L.
et al.: 1997b,
{\it Adv. Space Res.}, 20, 3. 

\bibitem[\protect\astroncite{Kohl et al.}{1998}]{koh98} 
Kohl, J.L.
et al.: 1998,
{\it \apj}, 501, L127.

\bibitem[\protect\astroncite{Li et al.}{1998}]{li98} 
Li, X., Habbal, S.R., Kohl, J.L. and  Noci, G.: 1998,
{\it \apj}, 501, L133.

\bibitem[\protect\astroncite{Noci, Kohl and Withbroe}{1987}]{noc87} 
Noci G., Kohl, J.L. and Withbroe, G.L.: 1987, 
{\it \apj}, 315, 706. 

\bibitem[\protect\astroncite{Noci et al.}{1997}]{noc97} 
Noci, G.
et al.: 1997,
{\it Adv. Space Res.}, 20, 2219. 

\bibitem[\protect\astroncite{Ofman  et al.}{1999}]{ofm99} 
Ofman, L., Nakariakov, V.M., DeForest, C.E.: 1999,
{\it \apj}, 514, 441.

\bibitem[\protect\astroncite{Peter and Judge}{1999}]{pet99}
Pete, H. and Judge, P.G. : 1999,
{\it \apj}, 522, 1148. 

\bibitem[\protect\astroncite{Strachan et al.}{1993}]{str93}
Strachan, L.
et al.: 1993,
{\it \apj}, 412, 410. 

\bibitem[\protect\astroncite{Wang}{1994}]{wan94}
Wang, Y.--M.: 1994,
{\it \apjl}, 435, L153. 

\bibitem[\protect\astroncite{Wilhelm et al.}{1998}]{wil98} 
Wilhelm, K.
et al.: 1998,
{\it \apj}, 500, 1023.

\end{thebibliography}
\end{document}